\documentclass[twocolumn]{aastex631}
\usepackage[latin1]{inputenc}
\usepackage{amsmath}
\usepackage{amsfonts}
\usepackage{amssymb}
\usepackage{subfigure}
\usepackage{graphicx}
\usepackage{xcolor}

\usepackage{natbib}
\newcommand{\steady}[0]{\textit{16TI} }
\newcommand{\spikes}[0]{\textit{40sp\_down} }

\listofchanges
\begin{document}

\newcommand{\davide}[1]{\textcolor{red}{#1}}

\title{Photospheric Prompt Emission From Long Gamma Ray Burst Simulations -- III. X-ray Spectropolarimetry} 
\author[0000-0002-4299-2517]{Tyler Parsotan}
\affiliation{Astrophysics Science Division, NASA Goddard Space Flight Center, Greenbelt, MD 20771, USA}

\author[0000-0002-9190-662X]{Davide Lazzati}
\affiliation{Department of Physics, Oregon State University, 301 Weniger Hall, Corvallis, OR 97331, USA.}

\begin{abstract} 
While Gamma Ray Bursts (GRBs) have the potential to shed light on the astrophysics of jets, compact objects, and cosmology, a major set back in their use as probes of these phenomena stems from our incomplete knowledge surrounding their prompt emission. There are numerous models that can account for various observations of GRBs in the gamma-ray and X-ray energy ranges due to the flexibility in the number of parameters that can be tuned to increase agreement with data. Furthermore, these models lack predictive power that can test future spectropolarimetric observations of GRBs across the electromagnetic spectrum. In this work, we use the MCRaT radiative transfer code to calculate the X-ray spectropolarimetric signatures expected from the photospheric model for two unique hydrodynamic simulations of long GRBs. We make time-resolved and time-integrated comparisons between the X-ray and gamma-ray mock observations, shedding light on the information that can be obtained from X-ray prompt emission signatures. Our results show that the $T_{90}$ derived from the X-ray lightcurve is the best diagnostic for the time that the central engine is active. We also find that our simulations reproduce the observed characteristics of the Einstein Probe detected GRB240315C. Based on our simulations, we are also able to make predictions for future X-ray spectropolarimetric measurements. Our results show the importance of conducting global radiative transfer calculations of GRB jets to better contextualize the prompt emission observations and constrain the mechanisms that produce the prompt emission. 
\end{abstract} 


\section{Introduction}
Gamma-ray Bursts (GRBs) are some of the most energetic events observed. These events were first detected in the 1960's as short pulses of gamma-ray radiation \citep{first_grbs} which  has since been determined to be produced by the associated GRB jet during the prompt emission phase. These jets produce $\sim 10^{51}$ ergs of energy, when the opening angle of the associated jet is taken into account \citep{Kulkarni_GRB_energy, ghirlanda2004collimation}, during the prompt phase of the event. This phase of GRBs is the least understood and the physical processes that produce this phase of emission are still under investigation. 

There have been many measurements made to attempt to better understand and characterize the radiation mechanism that produces the prompt emission. 
Typically, the prompt emission phase is characterized by the $T_{90}$ quantity which is defined as the time to accumulate 5-95\% of the photons that are observed in a given GRB event \citep{Parsotan_review}. It has been shown that the distribution of GRB $T_{90}$ is bimodal due to the two general types of GRB progenitors \citep{kouveliotou1993identification}. Long GRBs (LGRBs) typically have  $T_{90}\gtrsim 2$ s and are associated with core-collapse supernovae \citep{hjorth2003_LGRB_SNe, grb_collapsar_model, grb_sn_connection} while Short GRBs have $T_{90}\lesssim 2$ s and are associated with the merger of compact objects {\citep{GW_NS_merger, grb_NS_merger_connection, lazzati2018_GRB170817_afterglow}}{\citep{Eichler_sgrb_from_ns_bh,Narayan_sgrb_from_ns_bh,Mochkovitch_sgrb_from_ns_bh}}. These measurements are normally made in the gamma-ray energy range where there have been thousands of GRBs detected. However, there are also a number of X-ray prompt emission detections made by other instruments. The Monitor of All-sky X-ray Image (MAXI) is one such instrument that measures GRB prompt emission in the range of 0.5-30 keV \citep{matsuoka2009maxi} and sends alerts out to the community through the GCN. These observations provide important constraints to compare model predictions to. 

Recently, there have been a number of X-ray GRB observations and studies which show the importance of studying the prompt emission in this energy range. The Imaging X-ray Polarimetry Explorer (IXPE), conducted observations of GRB 221009A, the BOAT \citep{swift_boat, fermi_boat}, and was able to place an X-ray polarization upper limit of $\lesssim 82\%$ on the prompt emission from 2-8 keV {\citep{negroixpe_boat}}. While this upper limit is not very constraining, future instruments that can make these polarization measurements can begin to place novel constraints on prompt emission models. Another important observation is that of the Einstein Probe GRB, denoted GRB240315C or EP240315A, which has shown that the $T_{90}$ measured in the X-ray band can be substantially different from what is observed at gamma-ray energies \citep{EP240315a}. Besides these new observations, there have been recent efforts to characterize the prompt emission of GRBs down to X-ray energies \citep{oganesyan2017xray, oganesyan2018xray}. \cite{oganesyan2017xray} and \cite{oganesyan2018xray} showed that there may be low energy spectral breaks in GRB prompt spectra, with photon indexes that are different from what is seen in GRB spectra in the typical gamma-ray energy range. These observations and spectral studies motivate the production of falsifiable spectropolarimetric predictions at X-ray energies using GRB prompt emission models.

{There are two primary classes of GRB prompt emission models. Some models rely on synchrotron emission, such as the synchrotron shock model \citep{SSM_REES_MES} and the ICMART model \citep{ICMART_Zhang_2010}, while others rely on photospheric emission from the jet \citep{REES_MES_dissipative_photosphere, Peer_fuzzy_photosphere, Peer_photospheric_non-thermal} and dissipative processes which occur within the jet \citep{Atul, ito_mc_shocks, filip_rms}.}{There are many classes of models that can describe the prompt emission of GRBs. These models differ both in the location in which the prompt emission is produced and the types of processes which produce the emission. The jet's kinetic energy can be dissipated above the photosphere, which is the basis of the internal shock model \citep{SSM_REES_MES}, or near, including at, the photosphere, where there can be many processes that convert the dissipated energy to photons \citep{REES_MES_dissipative_photosphere} such as collisional heating \citep{Belo_collisional_photospheric_heating} and subphotospheric shocks \citep{ito_mc_shocks, filip_rms}. Furthermore, with the inclusion of magnetic fields, there can be magnetic dissipation occurring either above the photosphere, such as in the ICMART model \citep{ICMART_Zhang_2010}, or below, such as the case of magnetized radiation mediated shocks \citep{Belo_shocks}. } 

There are a number of advantages and disadvantages associated with each model however modelling efforts have historically focused on reproducing the wealth of GRB gamma-ray observations. Recently, in companion papers associated with this one, \cite{parsotan2021optical} and \cite{parsotan2022optical_pol} used the MCRaT radiative transfer code on a set of special relativistic hydrodynamic (SRHD) simulations of LGRB jets to produce optical spectropolarimetric predictions for GRBs which expand the testability of the non-dissipative photospheric model. Here, we expand on these associated works and produce prompt X-ray spectropolarimetric predictions from the same global radiative transfer MCRaT simulations. These collections of investigations allow for an unprecendented understanding of GRB broadband prompt spectropolarimetric signatures that provide both a new level of insight into the jet physics and a new level of predictions that can be tested with current and future observations. 

In this final paper of the series, we discuss the MCRaT code and the X-ray mock observations that
are constructed from the MCRaT simulations in Section \ref{global_methods}. In
Section 3 we show our results of the mock observed light
curves and polarizations for the set of LGRB
simulations analyzed with MCRaT. In Section 4, we
summarize our results and present them in the context of
current and future X-ray observations and the implications for the photospheric model and the underlying GRB jet physics.

\section{Methods}
\label{global_methods}
Many of the details of the simulations and the mock observations can be found in \cite{parsotan2021optical} and \cite{parsotan2022optical_pol}. We briefly summarize the relevant methods here.\footnote{The relevant codes, MCRaT data, and FLASH data for this paper can be found here: https://zenodo.org/records/11623566 \citep{parsotan_2024_dataset_codes}}

\subsection{The MCRaT Code}
The MCRaT radiative transfer code\footnote{The MCRaT code is open-source and is available to download at: https://
github.com/lazzati-astro/MCRaT/.} \citep{parsotan_mcrat_software_2021_4924630} allows for time-dependent radiative transfer calculations to be conducted using GRB jet profiles taken from FLASH \citep{fryxell2000flash} or PLUTO \citep{pluto_1, pluto_amr} SRHD GRB jet simulations. 

MCRaT operates by injecting photons into the SRHD simulation and Compton scattering them based on the local properties of the fluid. The code takes the full Klein-Nishina cross section into account, including polarization, and it takes cyclo-synchrotron emission and absorption into account as well \citep{parsotan_polarization, parsotan2021optical}.

\subsection{Mock Observations}
The lightcurve, spectra, and polarization mock observables are constructed using the ProcessMCRaT code\footnote{The code used to conduct the mock observations is also open-source and is
available at: https://github.com/parsotat/ProcessMCRaT} \citep{parsotan_processmcrat_software_2021_4918108}. The methods to construct these values and overviews of the spectral fittings that are done are outlined in \cite{parsotan_mcrat}, \cite{ parsotan_var},  \cite{parsotan_polarization}, and \cite{parsotan2021optical}. The spectral fittings provide the Band and COMP spectral parameters of $\alpha$, $\beta$, break energy, $E_o$, and peak energy, $E_\mathrm{pk}$. 

We can also calculate the time lag, $\tau$, between two lightcurves of different energy ranges. As was highlighted in \cite{parsotan2021optical}, this value is obtained by finding the time shift between the two lightcurves such that their spearman rank coefficient, $r_s$ is maximized. Here, we additionally calculate the $T_{90}$ of the lightcurves that we produce. The $T_{90}$ is calculated as $T_{95}-T_{5}$, where $T_{95}$ and $T_{5}$ are the times it takes to accumulate 95\% and 5\% of the photon counts in the lightcurve, respectively. These quantities are calculated based on the count rate of the lightcurve after it has been binned using the bayesian blocks algorithm \citep{astropy:2013, astropy:2018}. The counts in each timebin are cumulatively added and normalized by the total counts. Once the cumulative distribution of counts is calculated, we linearly interpolate the distribution to obtain the time when 5\% and 95\% of the counts have been accumulated. 

In this work, we focus on the X-ray prompt emission mock observables and making comparisons to the gamma-ray mock observables. The gamma-ray energy range is identical to that used in \cite{parsotan2022optical_pol} from $20-800$ keV, which corresponds to the POLAR-2 energy range \citep{polar_detector}. The gamma-ray spectral fittings are done in the Fermi GBM energy range of 8 keV and 40 MeV \citep{Yu_Bayesian_GRB_spectra}. The X-ray lightcurves, polarization degrees, $\Pi$, and polarization angles, $\chi$, are constructed in the energy range of $0.2 - 30$ keV, which primarily corresponds to the MAXI energy range \citep{matsuoka2009maxi} and includes the energy ranges encompassed by Swift XRT (0.2-10 keV; \cite{burrows2005swiftxrt}), NICER (0.2-12 keV; \cite{gendreau2016nicer}), IXPE (2-8 keV; \cite{weisskopf2016ixpe}), and Einstein Probe (0.5-8 keV; \cite{yuan2018einsteinprobe}). Furthermore, this relatively large energy range also allows for minimizing Poisson errors in the X-ray mock observables that are calculated from the MCRaT simulations.

Similar to prior works, we relate the mock observables to the structure of the simulated GRB jet through the use of equal arrival time surfaces (EATS) of the SRHD jet simulation. These EATS represent the location with the jet that would be emitting photons along a given observer's line of sight for a given time interval of the lightcurve \citep{parsotan_polarization, parsotan2022optical_pol}. 

\subsection{The Simulation Set}
The simulations presented here are identical to those presented in \cite{parsotan2021optical} and \cite{parsotan2022optical_pol}.

We used MCRaT to simulate the prompt emission from two FLASH SRHD LGRB jet simulations. Each jet was injected into a 16TI progenitor star \citep{Woosley_Heger}. One simulation injects a jet with constant luminosity, which we denote the \steady simulation. The jet is on for 100 s and is injected from a radius of $1 \times 10^9$ cm with an internal over rest mass ratio, $\eta=80$, an initial Lorentz factor of 5, and an opening angle of $10^\circ$ \citep{lazzati_photopshere}. The other simulation we denote the \spikes due to the jet being injected with pulses that are 0.5 seconds long with each pulse followed by another 0.5 seconds of quiescence. Each pulse of the injected jet luminosity is decreased by 5\% with respect to the initial pulse \citep{diego_lazzati_variable_grb}. In total the jet is active for 40 seconds before it is turned off, while the time in which it is actually on is $t_\mathrm{jet,on}=20$ s. All other characteristics of the injected \spikes jet is identical to that of the \steady simulation. The domain of the \spikes simulation is $2.56 \times 10^{12}$ cm along the jet axis while that of the \steady simulation is $2.5 \times 10^{13}$ cm along the jet axis. The resolution near the photosphere in each simulation is $\sim 10^9$ cm, and the temporal resolution of the FLASH simulation frames that are fed into MCRaT is 5 and 10 fps for the \steady and \spikes simulations, respectively. These spatial and temporal resolutions have been shown to be adequate for conducting MCRaT simulations with minimal errors introduced by the SRHD simulation \citep{arita-escalante2023}.

The MCRaT simulations are the same as those used in \cite{parsotan2021optical} and \cite{parsotan2022optical_pol}, where cyclo-synchrotron emission and absorption was taken into account.  Identical to these works, the MCRaT mock observations were constructed for a range of observer viewing angles, $\theta_\mathrm{v}$. For the \steady simulation, $\theta_\mathrm{v}$ ranges from $1-15^\circ$ while for the \spikes simulation the range is  $1-9^\circ$. {The spectra that we obtain here are also identical to those presented by \cite{parsotan2021optical} and \cite{parsotan2022optical_pol}, where the optical portion of the spectrum is due to the inclusion of cyclo-synchrotron emission and absorption while the high energy tail of the spectra are formed by the upscattering of photons in the photospheric region of the jet \cite{parsotan_mcrat} and bulk comptonization of the photons in the jet \cite{parsotan_var}.} Additional details can be found in these references.

\section{Results} \label{results}
In this section we outline our results from the X-ray spectropolarimetry analysis of the MCRaT LGRB simulations. First, we look at a time resolved analysis of the light curves and polarizations at X-ray and gamma-ray energies. Similar to prior studies, we relate these quantities to the locations of the X-ray and gamma-ray photons in the jet to get a better understand of the jet properties that are probed in each energy range. After that, we show the time-integrated polarization as a function of energy which can be directly compared to GRB observations. We primarily focus on the X-ray spectropolarimetric behavior but make relations to the gamma-ray spectropolarimetric results obtained by \cite{parsotan2022optical_pol}.

\subsection{Time Resolved Analysis}

\begin{figure*}[]
 \centering
 \gridline{
 \fig{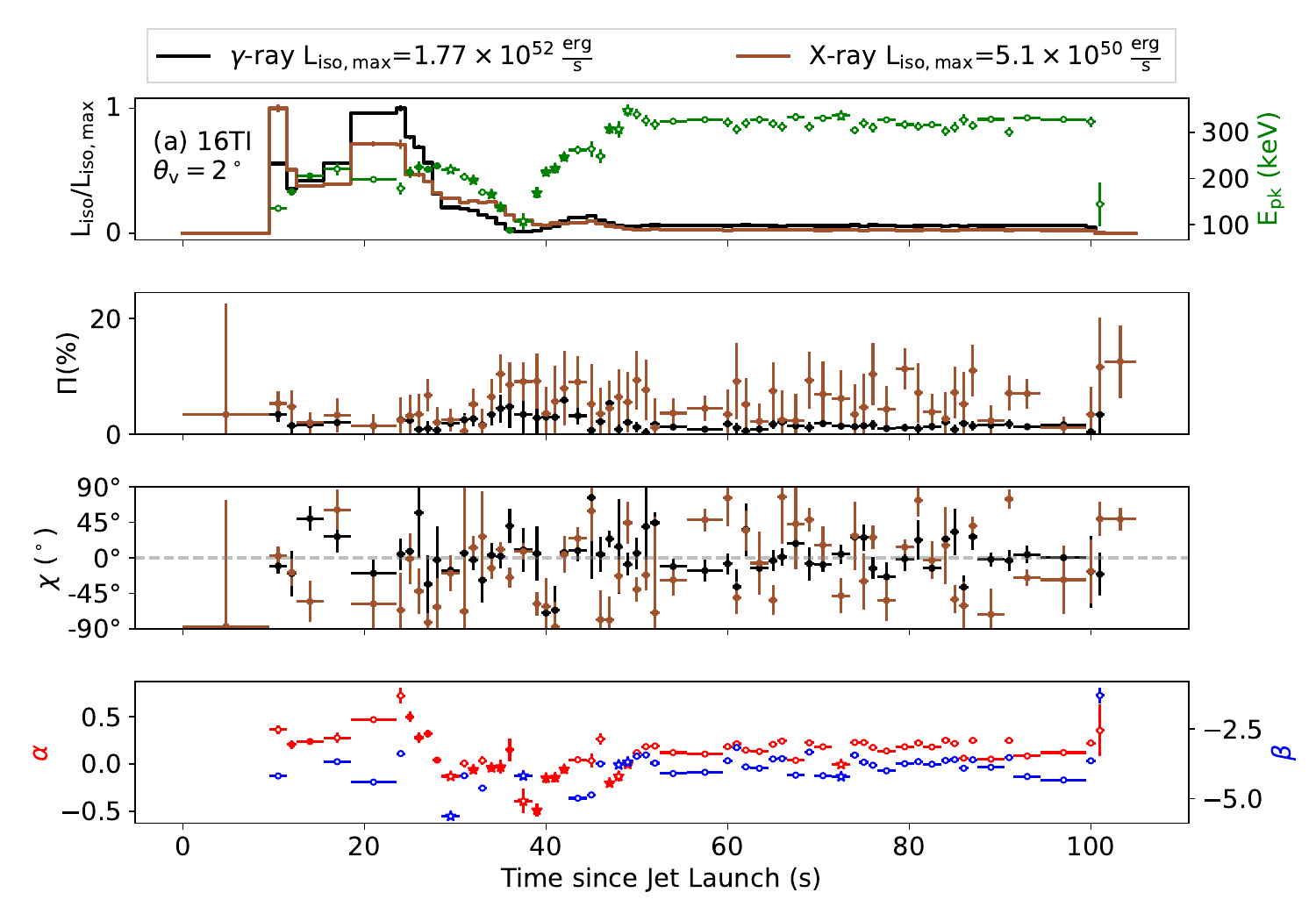}{0.8\textwidth}{\label{16ti_2_lc}}
 }
 \gridline{
 \fig{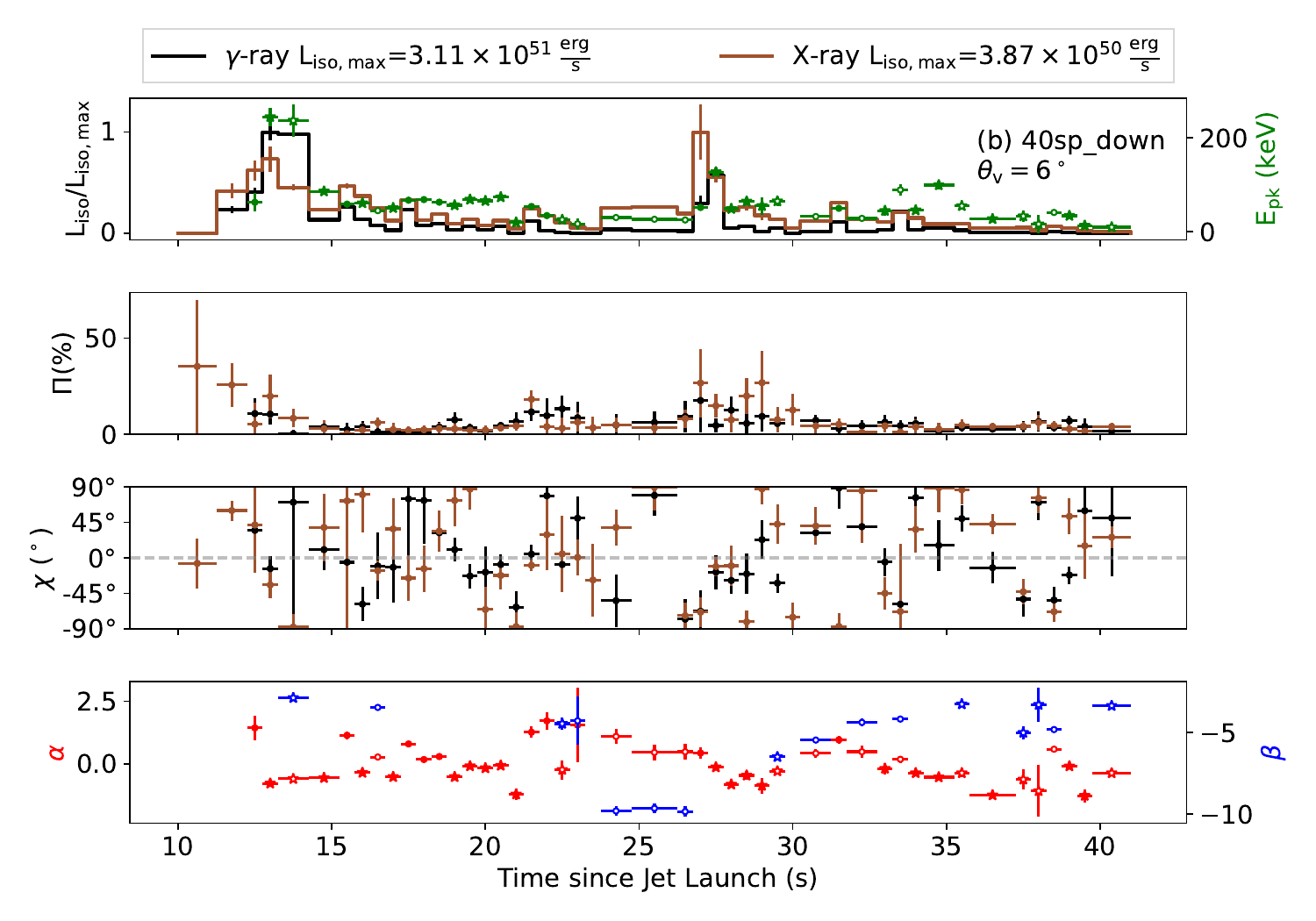}{0.8\textwidth}{\label{40sp_down_6_lc}}
 }
 \caption{Time-resolved mock observed quantities for the \steady simulation, for $\theta_\mathrm{v}=2^\circ$, and the \spikes simulation, for $\theta_\mathrm{v}=6^\circ$. In the top panels, we plot the time resolved fitted spectral peak energy E$_\mathrm{pk}$ in green and the gamma-ray and optical light curves in black and brown respectively, where each light curve is normalized by its own maximum value, L$_\mathrm{iso, max}$. The second panel shows the polarization degree of the gamma-ray and X-ray photons in black and brown. In the third panel, we show the polarization angles of the mock observed gamma-ray and X-ray photons. The dashed gray line denotes $\chi=0^\circ$. The last panel shows the time resolved spectral fitted $\alpha$ and $\beta$ parameters in red and blue respectively. The $\alpha$, $\beta$, and E$_\mathrm{pk}$ markers can be filled, denoting that the best fit spectrum is a COMP function, or unfilled to show that the Band function provides a superior fit. Any star $\alpha$, $\beta$, and E$_\mathrm{pk}$ markers, show spectra that are best fit with a negative $\alpha$ parameter.} 
 \label{light_curves}
\end{figure*}

Two examples of the time-resolved mock observed quantities for the \steady and \spikes simulations are shown in Figures \ref{light_curves}(a) and \ref{light_curves}(b) respectively. The mock observations are constructed for an observer located at $\theta_\mathrm{v}=2^\circ$ with respect to the jet axis of the \steady simulation, in Figure \ref{light_curves}(a), and an observer located at $\theta_\mathrm{v}=6^\circ$, for the \spikes simulation, shown in Figure \ref{light_curves}(b). In the top panel of each figure, we show the gamma-ray lightcurve in black normalized by it's maximum luminosity, $L_\mathrm{iso, max}$, while the X-ray lightcurve normalized by it's own maximum is shown in brown. Additionally, the time resolved spectral $E_\mathrm{pk}$ are shown in green. The second panels show the polarization degree, $\Pi$, for the gamma-ray and X-ray energy ranges using the same color scheme. The third panels show the polarization angle, $\chi$, using the same colors for each energy range as well as a dashed grey line denoting  $\chi=0^\circ$. The final panel shows
the fitted $\alpha$ and $\beta$ parameters, based on the type of fit, in red and blue respectively; unfilled $\alpha$, $\beta$, and $E_\mathrm{pk}$ markers represent spectra that are best fit by a Band function and filled, solid markers represent spectra best fit with the COMP function. Additionally, star markers identify spectra where $\alpha < 0$. 

We find that the time-resolved X-ray polarization is relatively low, peaking at $\sim 25\%$ in certain time intervals but consistently staying below $\sim 10\%$ for many time intervals. 

\begin{figure*}
    \centering
    \includegraphics[width=\textwidth]{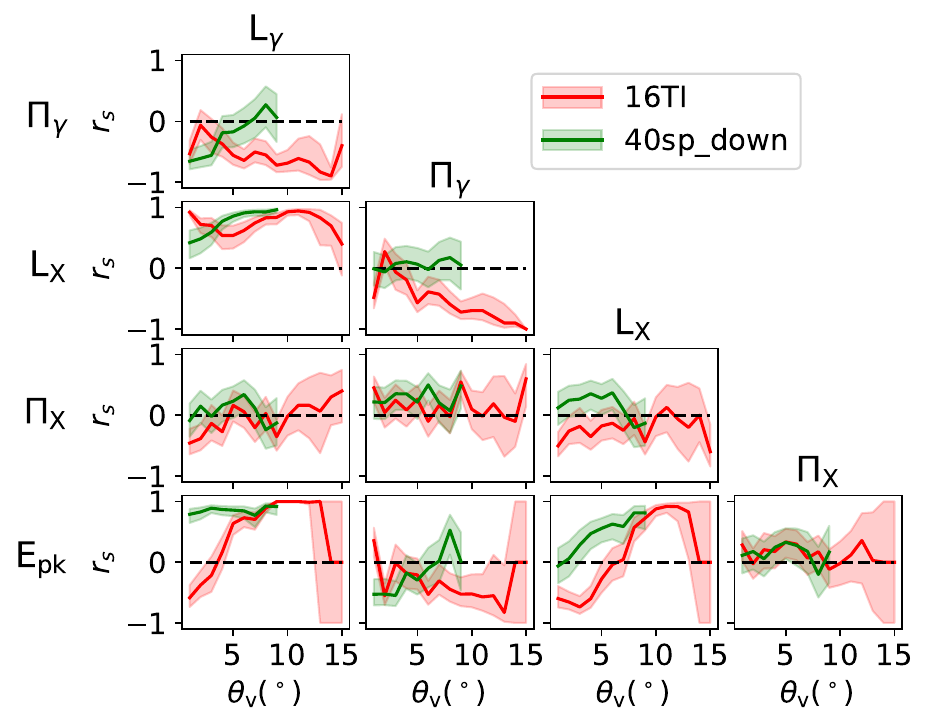}
    \caption{The spearman rank coefficient, $r_s$, between various time-resolved mock observed quantities for the \steady and \spikes simulations, shown in red and green respectively, as a function of $\theta_\mathrm{v}$. The 95\% confidence interval regions are shown as shaded regions. The column and row titles of a given plot show the quantities that were used to calculate $r_s$ at a given $\theta_\mathrm{v}$. As a note, the confidence intervals for comparisons with $E_\mathrm{pk}$ get large for the \steady simulation at $\theta_\mathrm{v} \gtrsim 11^\circ$ due to the low number of time-resolved spectra with fits that are well constrained.
    }
    \label{fig:correlations}
\end{figure*}

To understand the tracking behavior between the different time resolved observables, we calculate the spearman rank coefficient, $r_s$ using different permutations of the mock observed quantities. These are shown in Figure \ref{fig:correlations}, where we plot $r_s$ for different quantities of the \steady and \spikes simulations in red and green respectively including the 95\% confidence region. Figure \ref{fig:correlations} shows a few notable behaviors between the X-ray mock observables and a few others. One is the lack of any correlation, either positive or negative, between the X-ray polarization degree and any other mock observable quantity. Another behavior that we see is between the X-ray lightcurve and the spectral $E_\mathrm{pk}$ there the $r_s$ evolves from a negative, or no, correlation to a positive correlation as $\theta_\mathrm{v}$ increases in both the \steady and \spikes simulations. For both simulations, we also see that the X-ray lightcurve is positively correlated with the gamma-ray lightcurve. Finally, we find that the X-ray lightcurve seems to be inversely correlated with the gamma-ray polarization degree particularly in the \steady simulation at large $\theta_\mathrm{v}$. 
Similarly, as was previously obtained by \cite{parsotan2022optical_pol}, $\Pi_\gamma$ and $L_\gamma$ are also inversely correlated.

\begin{figure}
    \centering
    \includegraphics[width=0.5\textwidth]{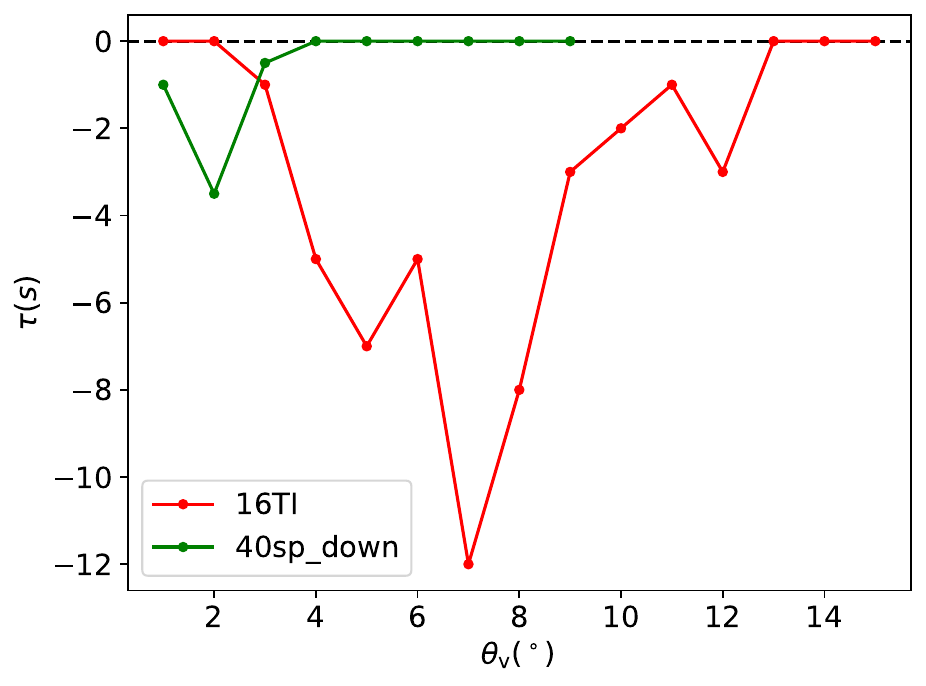}
    \caption{The time lag between the gamma-ray and X-ray lightcurves as a function of observer viewing angle for the \steady and \spikes simulations, shown in red and green respectively. The gamma-ray and X-ray lightcurves for the \spikes simulations experience little to no lag while the X-ray lightcurve peak can precede the gamma-ray lightcurve peak by $\sim 10$ s for the \steady simulation. }
    \label{fig:xray_gamma_delay}
\end{figure}

Focusing further on the lightcurves, we see that the correlation between the X-ray and gamma-ray lightcurves is also shown in Figure \ref{fig:xray_gamma_delay}, where we plot the time lag, $\tau$, between the gamma-ray and the X-ray lightcurves as a function of $\theta_\mathrm{v}$. This quantity for the \spikes simulation, shown in green, is $\lesssim 4$ s, which shows that there is very little lag and the peaks of the lightcurves in each energy range are closely aligned to one another. The negative value of $\tau$ indicates that the X-ray peak does precedes the gamma-ray peak by $\tau \lesssim 4$ s. The lag for the \steady simulation, shown in red, can be as large as $\sim 10$ s, where the X-ray peak once again precedes the gamma-ray peak by this amount.

\begin{figure}
    \centering
    \includegraphics[width=0.5\textwidth]{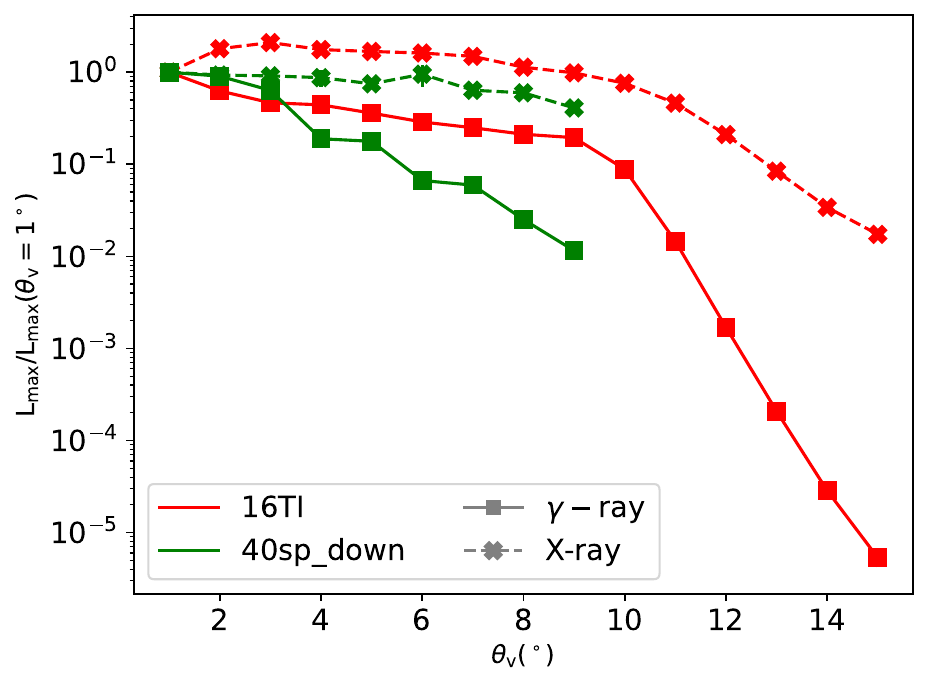}
    \caption{The maximum gamma-ray and X-ray lightcurve isotropic luminosities of the \steady and \spikes simulations as a function of observer viewing angle. Luminosities are normalized to the maximum luminosity of an observer located at $\theta_\mathrm{v}=1^\circ$. The \steady simulation's gamma-ray and x-ray values are plotted in red as solid and dashed lines, with square and X marker respectively. The \spikes simulation data points and lines are plotted in green. This figure shows that the X-ray peak luminosity decays very slowly as a function of $\theta_\mathrm{v}$ while the peak gamma-ray luminosity can be many orders of magnitude dimmer than what it would be on axis.}
    \label{fig:peak_lumi_angle}
\end{figure}

We characterize how the maximum luminosity of the X-ray and gamma-ray lightcurves change as a function of viewing angle to understand how bright the signals may be, especially when observing an off-axis event. Figure \ref{fig:peak_lumi_angle} shows the maximum isotropic lightcurve luminosity for the \steady and \spikes simulations in red and green respectively. The gamma-ray and X-ray isotropic $L_\mathrm{max}$ are shown as solid and dashed lines with square and X markers respectively. We find that the gamma-ray $L_\mathrm{max}$ decreases rapidly compared to the X-ray $L_\mathrm{max}$ for all $\theta_\mathrm{v}$ in both simulations. This result is in line with the fact that the jet Lorentz factor in each simulation decreases moving from the core to high latitude regions of the jet (see Figure 3 of \cite{parsotan_polarization}) which also causes the spectra to become more X-ray photon dominated at large $\theta_\mathrm{v}$. 

\begin{figure}
    \centering
    \includegraphics[width=0.5\textwidth]{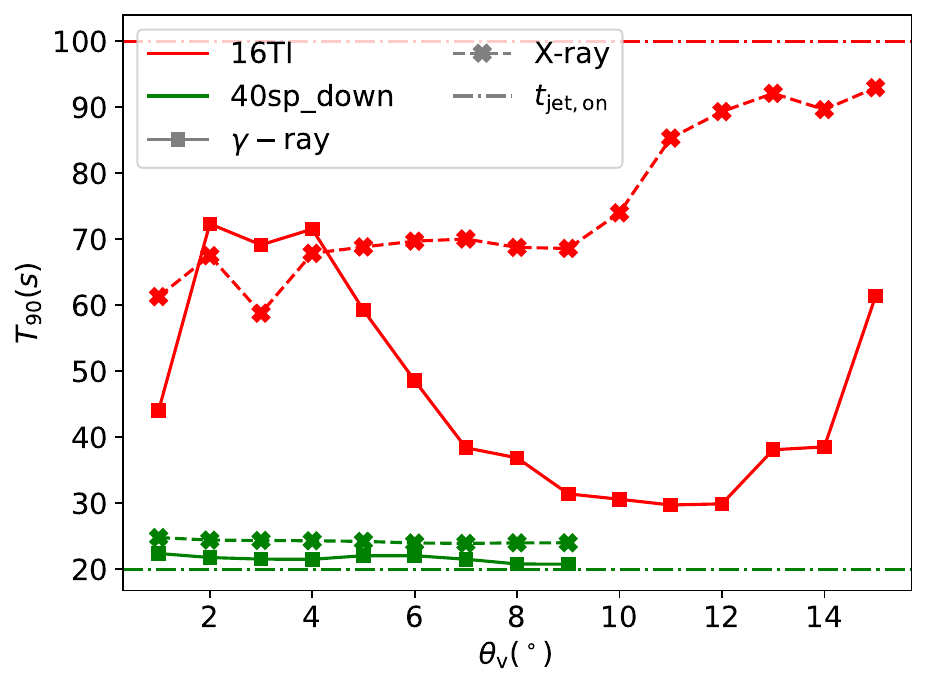}
    \caption{The calculated $T_{90}$ of the gamma-ray and X-ray lightcurves as a function of observer viewing angle. The $T_{90}$ for the \steady simulation is shown in red, where the values measured from the gamma-ray lightcurves are shown as a solid line with square markers while the X-ray lightcurve derived values are plotted with a dashed line with X markers. The same quantities are plotted in green for the \spikes simulation. The total time that the jet is on in each simulation, $t_\mathrm{jet,on}$, is shown as a dash-dotted line in red and green, for the \steady and \spikes simulations respectively. The X-ray $T_{90}$ in the \steady simulation more closely approximates the time that the jet is on, especially as $\theta_\mathrm{v}$ increases. Generally, the X-ray emission lasts longer than the gamma-ray emission in both simulations.}
    \label{fig:t90}
\end{figure}

In addition to focusing on the maximum luminosities of the X-ray and gamma-ray lightcurves, we can also look at the duration of the emission in those energy bands using $T_{90}$. Figure \ref{fig:t90} shows the calculated $T_{90}$ for the \steady and \spikes simulations in red and green, respectively. The gamma-ray calculated $T_{90}$ are shown as a solid line with square markers while the X-ray $T_{90}$ are denoted with dashed lines and X markers. We additionally show the time that that jet is on, $t_\mathrm{jet,on}$, for the simulations with dashed-dotted lines. We find that in the case of the \spikes simulation, the  $T_{90}$ are relatively consistent between the gamma-ray and X-ray measured quantities. Furthermore, these $T_{90}$ are very close to the total time that the jet is on in the \spikes simulation, where $t_\mathrm{jet,on}=20$ s. On the other hand, looking at the \steady simulation, we can see that the gamma-ray derived $T_{90}$ can be significantly different from the X-ray measured value by up to a factor of $\sim 3$. We also find that in the case of the \steady simulation, the measured X-ray $T_{90}$ more closely approaches the total time that the \steady simulation jet is on, $t_\mathrm{jet,on}=100$ s, for observers that are located off-axis. Part of the difference in the $T_{90}$ values is related to the presence of X-ray precursors that we find in the \steady simulation particularly for relatively large $\theta_\mathrm{v}$. These precursors occur up to $\sim 10$'s of seconds before the main X-ray and gamma-ray emission.  

\begin{figure*}
\begin{interactive}{animation}{16TI_test_7.mp4}
\plotone{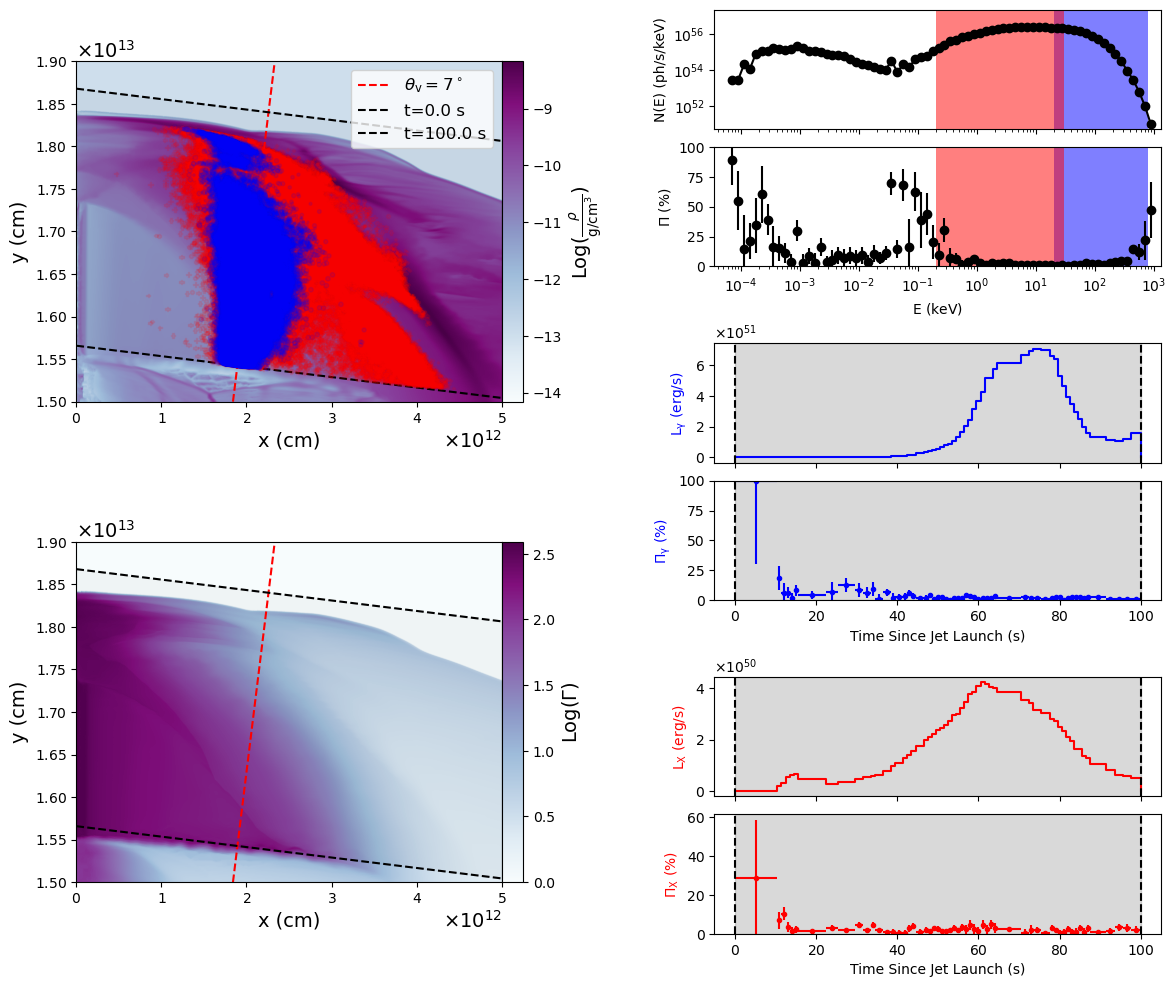}
\end{interactive}
\caption{The relation between the mock observed quantities and the jet structure of the \steady simulation for $\theta_\mathrm{v}=7^\circ$. The top left panel shows a pseudo-color logarithm density plot of the simulated jet. The two dashed black lines denote the EATS for the times specified in the legend, and the red dashed line shows the line of sight of the observer from infinity to the central engine of the simulated GRB. Also shown on the pseudo-color plot are the detected X-ray and gamma-ray photons, in red and blue translucent markers respectively, for the time period shown in the legend. These photon markers show where the photons are located in the jet. Additionally, the markers are translucent allowing us to identify regions of the jet where the photons are densely located (due to the concentration of blue or red), and the markers are different sizes to show the weight of each photon in the calculation of the various mock observable quantities. The bottom left panel shows a pseudo-color plot of the simulated jet's bulk Lorentz factor with the same black dashed lines that are plotted in the top left panel. The top right panels show the spectrum in units of counts and the energy resolved polarization degree, for the time interval highlighted in the pseudo-color density plot. The red and blue highlighted regions show the energy ranges that are used to calculate the X-ray and gamma-ray mock observables, respectively. The bottom right four panels show these mock observed quantities -- the gamma-ray light curve and time resolved polarization in blue in the middle two panels and the X-ray quantities in red in the bottom two panels. In each lightcurve and polarization panel there are black dashed lines that shows the time interval of interest which correspond to the plotted photons in the pseudo-color density plot.
This figure is available as {a 19 second long} animation where it steps through time intervals in the light curves and plots the location of the X-ray and gamma-ray photons, in relation to the jet structure, as well as the spectrum of those same photons. {The location of the photons in the jet changes for each time interval of the mock observed lightcurve. We see both the gamma-ray and X-ray photons located primarily along the observers line of sight as we look at each time slice of mock observed photons.}
}
\label{16ti_ani}
\end{figure*}

\begin{figure*}
\begin{interactive}{animation}{40sp_down_test_2.mp4}
\plotone{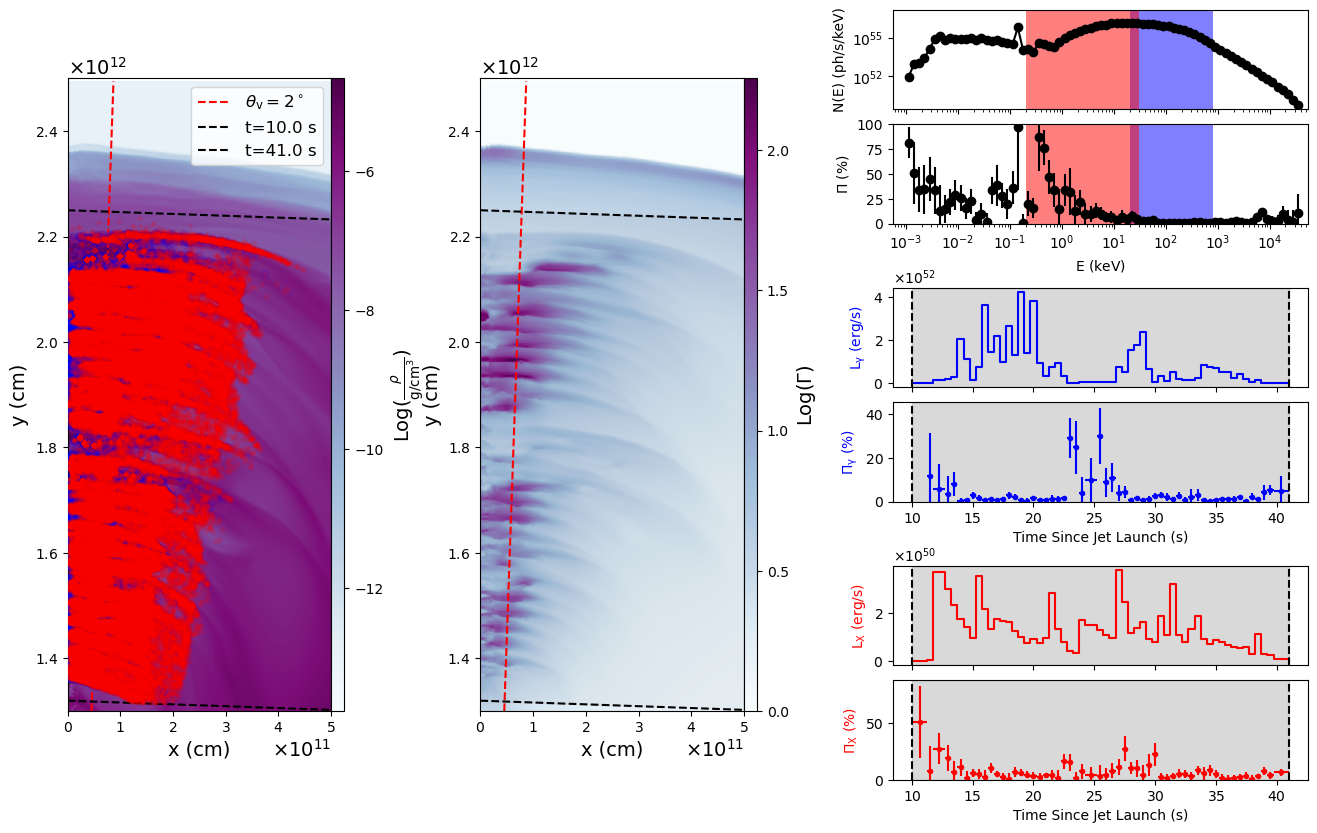}
\end{interactive}
\caption{The relation between the mock observed quantities and the jet structure of the \spikes simulation for $\theta_\mathrm{v}=2^\circ$. The formatting is identical to Figure \ref{16ti_ani}, with the exception of the placement of the panels. The left most panel shows the pseudo-color density plot with the locations of the gamma-ray and X-ray photons plotted while the middle panel shows the bulk Lorentz factor of the jet. The top right panels show the spectrum of the photons and the energy resolved polarization. The middle panels show L$_\gamma$ and $\Pi_\gamma$ while the bottom right panels show L$_\mathrm{X}$ and $\Pi_\mathrm{X}$. This figure is also available as {a 19 second long} animation. {The location of the photons in the jet changes for each time interval of the mock observed lightcurve. Similar to the \steady simulation we see both the gamma-ray and X-ray photons located primarily along the observers line of sight as we look at each time slice of mock observed photons. The X-ray photons better trace out the shells of material that are produced by the variable injected jet and are less sensitive to the bulk lorentz factor of the fluid allowing the X-ray emission to persist when the gamma-ray lightcurve is quiescent.}}
\label{40sp_down_ani}
\end{figure*}

To understand the precursor and associated $T_{90}$ behavior, we plot the location of the gamma-ray and X-ray photons in relation to the SRHD jet structure. These are shown in Figure \ref{16ti_ani} and Figure \ref{40sp_down_ani}, which are also available as animations. Figure \ref{16ti_ani} shows the locations of the photons in relation to the jet profile for the  \steady simulation where the photons are those that would be detected by an observer located at $\theta_\mathrm{v}=7^\circ$. In the top left panel of the figure we show a pseudo-color density plot of the \steady simulation. Overlaid are the gamma-ray photons and the X-ray photons in blue and red respectively, where the markers are translucent and the marker sizes represent the weight of the photon. In the bottom left panel of the figure, we plot the pseudo-color Lorentz factor, $\Gamma$, profile of the jet which provides information on the $1/\Gamma$ scattering angle of photons. On the right hand side of the panel, we plot the mock observable quantities of the photon count spectrum and polarization degree as a function of energy, the gamma-ray lightcurve and polarization degree as a function of time, and finally the X-ray lightcurve and polarization degree as a function of energy. For the X-ray lightcurve presented here, there is precursor X-ray emission prior to the peak of the X-ray and gamma-ray lightcurves. We can see, particularly in the animated version of the plot, that the precursor photons originate from near the core of the jet. Then, as the observer sees fluid that is directly boosted towards them, they see a peak in the gamma-ray and X-ray lightcurves at the same time since the photons are co-located in the jet. The X-ray emission from the dense jet cocoon interface does not contribute much as the polarization degree is very low over the course of the lightcurve, indicative of the fact that the photons along the observer's line of sight contribute more to the measured signals.  

Figure \ref{40sp_down_ani} shows the same information as Figure \ref{16ti_ani}, for the \spikes simulation, with $\theta_\mathrm{v}=2^\circ$. Here, the pseudo-color density and $\Gamma$ plots are ordered left-to-right instead of top-to-bottom. For the \spikes simulation, we find that the X-ray photons more closely trace the shocked shells in the jet and there are many more pulses in the X-ray lightcurve compared to the gamma-ray lightcurve. In this simulation, the X-ray photons along the observer's line of sight primarily contribute to the emission that is measured, similar to the \steady simulation, however the X-ray photons are not as sensitive to the fluid $\Gamma$ which allows for X-ray emission to still be detected even when there is a dip in the gamma-ray lightcurve (see the difference in gamma-ray and X-ray lightcurve emission at $t\sim 25$ s since jet launch). 

\subsection{Time Integrated Analysis}

We also calculate the time-integrated quantities that can be used to compare to current and future observations. 

\begin{figure}
    \centering
    \includegraphics[width=0.5\textwidth]{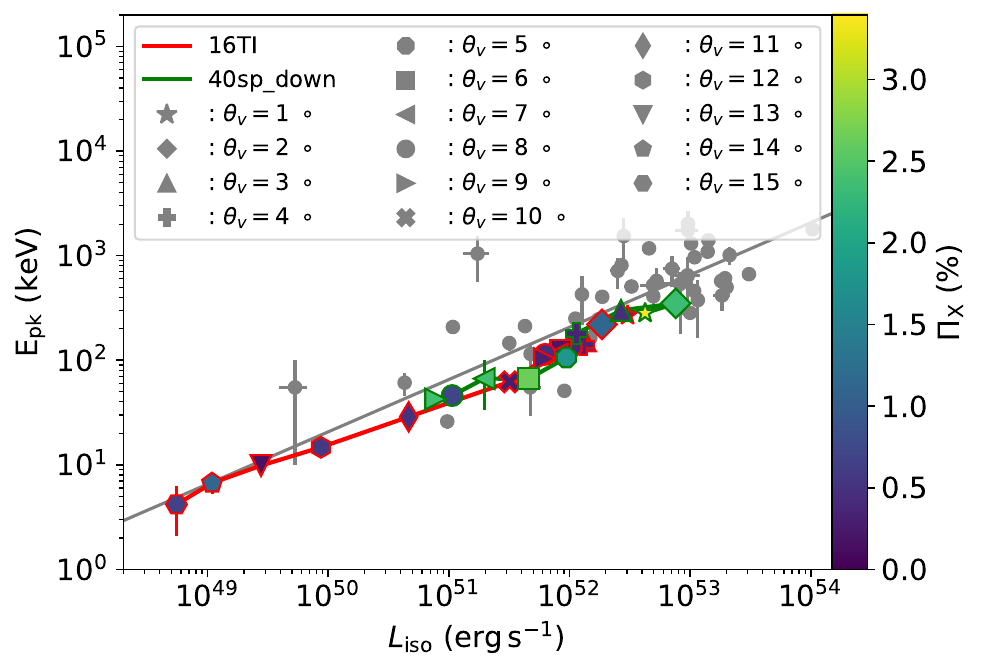}
    \caption{The location of the \steady and \spikes simulations alongside the Yonetoku relation \citep{Yonetoku}, plotted in red and green respectively. The different markers denote where an observer located at different $\theta_\mathrm{v}$ would place the observed simulation alongside the relation. The fill color of the markers denote the time-integrated X-ray polarization that the same observer would measure. The Yonetoku relation is shown as the gray solid line and observed data from \cite{data_set} are plotted as gray markers. }
    \label{fig:yonetoku}
\end{figure}

In Figure \ref{fig:yonetoku} we show the location of the \steady and \spikes simulations in relation to the Yonetoku relation \citep{Yonetoku}, in red and green respectively. The Yontetoku relation is shown as the gray line while observed GRBs from \cite{data_set} and their position along the relationship is shown as grey circle markers. The different marker shapes for the MCRaT simulation points denote the $\theta_\mathrm{v}$ for which the data point is obtained. We additionally fill the markers with colors that correspond to the time-integrated X-ray polarization that would be measured by the same observer located at $\theta_\mathrm{v}$. We find, analogous to prior results, that the MCRaT simulations are able to reproduce the slope and normalization of the Yonetoku relation. However, unlike prior analyses that looked at optical prompt polarization \citep{parsotan2022optical_pol},  the expected time-integrated polarization is very low where $\Pi_\mathrm{X} \lesssim 3\%$. In comparing the time-integrated $\Pi_\mathrm{X}$ to the time-integrated $\Pi_\gamma$ obtained by \cite{parsotan2022optical_pol}, we also find that the two exhibit slightly different behaviors. The time-integrated $\Pi_\mathrm{X}$ seem to decrease as a function of $\theta_\mathrm{v}$ while the time-integrated $\Pi_\gamma$ increases as $\theta_\mathrm{v}$ increases.

\begin{figure}
    \centering
    \includegraphics[width=0.5\textwidth]{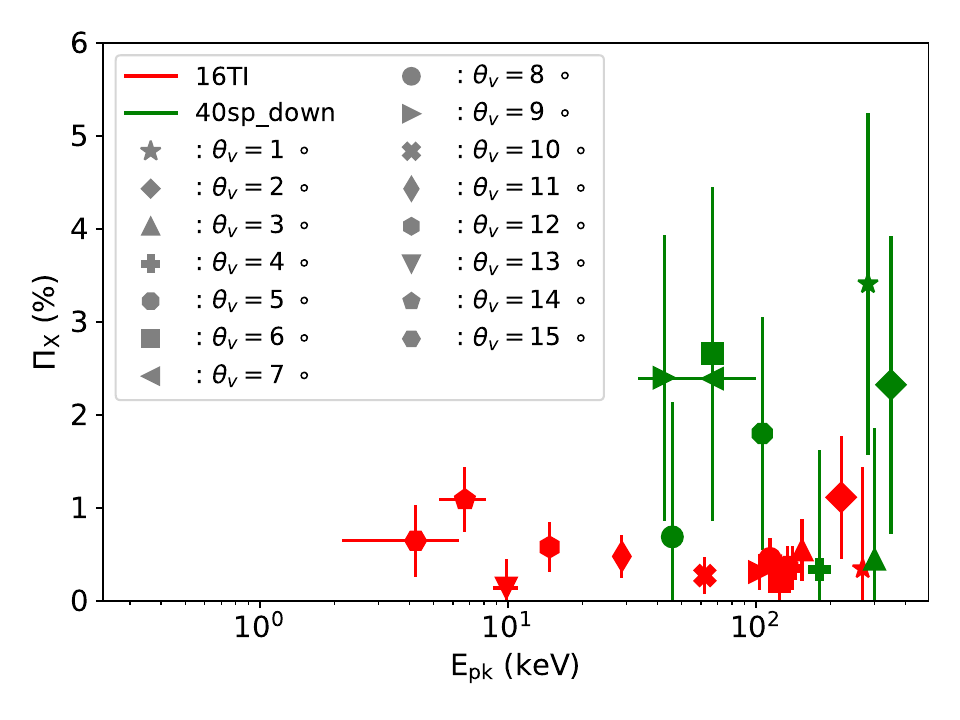}
    \caption{Time-integrated X-ray polarization as a function of time integrated spectral $E_\mathrm{pk}$ for the \steady and \spikes simulations, shown in red and green respectively. The different markers show the values that are obtained from observers located at different $\theta_\mathrm{v}$.}
    \label{fig:pol_energy}
\end{figure}

We also find that there is no evolution of the time-integrated X-ray polarization with the time integrated spectral $E_\mathrm{pk}$, which is shown in Figure \ref{fig:pol_energy}. The red and green points show the measured values for the \steady and \spikes simulations respectively. The different markers denote the values that would be measured for observers located at various $\theta_\mathrm{v}$. Regardless of $\theta_\mathrm{v}$, the time-integrated X-ray polarization degrees are all consistent with one another, within their $1\sigma$ errorbars. This Figure further shows the low time-integrated X-ray polarization values that were also shown in Figure \ref{fig:yonetoku}. 

\section{Summary and Discussion} \label{summary}
In this work, we have used the MCRaT simulations that were presented in companion papers \citep{parsotan2021optical, parsotan2022optical_pol} to investigate the prompt X-ray behavior of LGRBs expected in the photospheric model. The two LGRB jet simulations are denoted as the \steady simulation, where the jet is injected with a constant luminosity profile, and the \spikes simulation, where the jet luminosity profile is variable in time. The MCRaT simulations take cyclo-synchrotron emission and absorption into account in simulating the emission produced in these jets. The X-ray mock observations are constructed for the energy range of $0.2-30$ keV which corresponds to the MAXI energy range \citep{matsuoka2009maxi} and the gamma-ray energy range is defined to be $20-800$ keV, which is the energy range of the POLAR-2 polarimeter \citep{polar_detector}. The results of constructing the X-ray mock observables and comparing them to the gamma-ray quantities provide a physical basis for understanding what portion of the jet is being probed by each energy band. 

The main results from analyzing the simulated X-ray prompt emission from our set of SRHD simulations can be summarized as follows:
\begin{itemize}
    \item[1.] Both the time-resolved and time-integrated X-ray polarization degree, $\Pi_\mathrm{X}$, is expected to be very low with the time-resolved $\Pi_\mathrm{X} \lesssim 10\%$ and the time-integrated $\Pi_\mathrm{X} \lesssim 3\%$
    
    \item[2.] The X-ray emission generally tracks the gamma-ray emission, with the lightcurve peaks generally occuring within $\sim 10$ seconds of one another
    
    \item[3.] The X-ray emission generally lasts longer than the gamma-ray emission, and the $T_{90}$ measured from the X-ray emission is a better indicator of the time that the jet is active for, at least in the case of the \steady simulation. 
        
    \item[4.] Related to the longer X-ray $T_{90}$, is the presence of X-ray precursors that precede the main X-ray and gamma-ray lightcurve peaks  
    
    \item[5.] The peak luminosity of the X-ray lightcurve has a weak dependence on $\theta_\mathrm{v}$ while the peak luminosity of the gamma-ray lightcurve drops off sharply as a function of $\theta_\mathrm{v}$.
    
    \item[6.] The X-ray photons primarily probe the same material as the gamma-ray photons. However, the gamma-ray photons originate from the jet material that is located directly along the observers line of sight and are thus more sensitive to the bulk $\Gamma$ of the jet. The X-ray photons are less sensitive to the fluid properties directly along the observers line of sight which means that the emission is expected to be more stochastic and long-lasting. 
\end{itemize}

The low $\Pi_\mathrm{X}$ that we find in this work is in contrast to the values that have been previously obtained for optical and gamma-ray polarizations, $\Pi_\mathrm{opt}$ and $\Pi_\gamma$, respectively. We find that the time integrated $\Pi_\mathrm{X}$ is $\lesssim 3\%$ and it does not seem to change much as a function of $\theta_\mathrm{v}$. The time integrated $\Pi_\gamma$ from these same simulations is low when the observer is on-axis and increases as a function of $\theta_\mathrm{v}$ while time-integrated $\Pi_\mathrm{opt}$ can be relatively large when $\theta_\mathrm{v}$ is small and  decreases as $\theta_\mathrm{v}$ increases \citep{parsotan2022optical_pol}. The time-resolved $\Pi_\mathrm{X}$ generally remain very low throughout time in our simulations due to the X-ray photon emitting portion of the jet being primarily along the observer's line of sight. In the case of there being X-ray precursors, as is shown in Figure \ref{16ti_ani}, the X-ray emission is asymmetric about the observer's line of sight which leads to larger $\Pi_\mathrm{X}$ at these early times in the lightcurve. The general behavior that we see for the time-resolved $\Pi_\mathrm{X}$ is analagous to what was found for the time resolved $\Pi_\gamma$, and is reproduced here in this work \citep{parsotan2022optical_pol}. The time-resolved $\Pi_\mathrm{X}$, however, is in contrast to the large time-resolved $\Pi_\mathrm{opt}$ that has been shown to be possible from these simulations thorughout the lightcurve, due to the nearly constant asymmetry of the emitting region of the optical photons with respect to the observer's line of sight \citep{parsotan2022optical_pol}. 

Our combined results, of finding low $\Pi_\mathrm{X}$ and a positive correlation between the X-ray and gamma-ray lightcurves, are in line with the X-ray and gamma-ray photons being co-located in the outflow, primarily along the observers line of sight. The peak of the two lightcurves are generally simultaneous or within $\sim 10$ seconds of one another in our simulations, and the coincident peaks are due to the observer seeing photons probing the jet material that is Lorentz boosted directly towards the observer. The X-ray photons are not as dependent on the jet material being boosted towards the observer and thus the lightcurve emission can persist even when the gamma-ray emission is faint, due to the jet Lorentz factor being low. Overall, this finding lends itself to our result that the X-ray $T_{90}$ is a better indicator of the jet engine active time, as is shown in Figure \ref{fig:t90}. Furthermore, we find X-ray precursors that occur early in the lightcurve before the main peaks of the X-ray and gamma-ray emission. These precursor signals, if appropriately localized, can be used as a warning signals such that multiwavelength facilities can observe the localized region of the sky to conduct observations of the peak of the X-ray and gamma-ray prompt emission. These precursor signals would allow for the unprecedented opportunity to conduct ''follow-up`` campaigns of the prompt emission placing stringent constraints on GRB prompt emission models. This is particularly promising as the peak of the X-ray prompt emission depends very weakly on $\theta_\mathrm{v}$, as is shown in Figure \ref{fig:peak_lumi_angle}. {A full morphological study of the pulse profiles that we find in the MCRaT lightcurves is warranted but is outside the scope of this study. Future works will aim to fully characterize the morphologies of the lightcurves that we obtain from these simulations.}

Our findings are in line with the recent observations of GRB240315C/EP240315A \citep{EP240315a}. This GRB was first detected in the 0.5-4 keV energy range by the Wide-field X-ray Telescope (WXT) on board the Einstein Probe observatory. The emission in this energy band was highly variable and lasted for $\sim 1500$ s, with a measured $T_{90}$ of $\sim 1000$ s. The gamma-rays from this transient was detected by the Swift Burst Alert Telescope (BAT) and Konus-Wind $\sim 370$ s after the initial detection by WXT. {The gamma-ray emission is coincident with the peak of the X-ray emission.}{While the gamma-ray emission is coincident with the peak of the X-ray emission, \cite{EP240315a} found that if the source was in the BAT field of view early on it may have been detectable in Swift-BAT's image mode which makes interpreting the behavior of the gamma-ray lightcurve difficult.} Based on the localization of the initial pulse of X-ray emission detected by WXT, a redshift was able to be measured for this event as $z=4.859$. Based on the simulations and mock observations presented here, this GRB was most likely viewed off axis. The initial pulse of X-rays was produced by the core of the jet while the coincident gamma-ray and X-ray peaks in the lightcurve was produced by the jet material that is Lorentz boosted towards the observer. The subsequent X-ray emission is from the jet material that is moving at a Lorentz factor that is smaller than that correspondent to the peak of the X-ray and gamma-ray lightcurves. In this work, we find that the X-ray $T_{90}$ is a better indicator of the time that the jet engine is active which, for GRB240315C, comes out to be $t_\mathrm{jet,on}\sim T_{90,X}/(1+z)\sim 170$ s.

As was highlighted in the companion papers \citep{parsotan2021optical, parsotan2022optical_pol}, the mock observed spectra at low energies do not generally reproduce the Band $\alpha \sim -1$. Instead, the MCRaT spectra have $\alpha \sim 1$ which is a well known issue with the photospheric model. As a result of this, the luminosities and energies that we obtain from our mock X-ray observations are lower limits on what is truly expected in GRBs, while our results regarding $T_{90}$ should be independent of this caveat. This means that, particularly for off-axis GRBs, the X-ray prompt emission will be observable but the gamma-ray emission will not be present due to it's strong dependence on $\theta_\mathrm{v}$. In these cases, we would still expect there to be associated optical prompt emission \citep{parsotan2021optical}. To properly reproduce the low energy spectral properties of GRBs, we need to include a mechanism that will produce low energy photons into the outflow. There are not enough cyclo-synchrotron photons at low energies in the current suite of MCRaT simulations which indicates that synchrotron emission and absorption can play a dominant role at these energies. Other radiative mechanisms that can reproduce the $\alpha \sim -1$ values include radiation mediated shocks \citep{filip_rms, ito_mc_shocks}. Incorporating both synchrotron and radiation mediated shocks into the MCRaT code will be the focus of future works. 

Another caveat to our analysis is related to the extremely low $\Pi_\mathrm{X}$ that we obtain. These low values, while consistent with the upper limit placed on GRB 2210009A by IXPE \citep{negroixpe_boat}, are primarily due to the geometric effects about the observer's line of sight. Other works have looked at optically thin synchrotron in the dissipative photospheric model and found significant polarization of $\Pi_\mathrm{X}\sim 50\%$ at $\sim 0.3$ keV \citep{Lundman_polarization}. We expect that adding synchrotron emission and absorption physics into MCRaT will not only allow for the mock observed spectra to come into alignment with spectra from observations, but will also align with these results and cause the expected $\Pi_\mathrm{X}$ from realistically structured jets to also significantly increase. 

The optical to gamma-ray spectropolarimetric analyses presented here, and in the companion papers to this one, have shed light on the broadband prompt emission expected from the photospheric model. We are able to gain new insights into the physical characteristics of the jets that each energy range can provide, which gives us a basis for interpreting a myraid of observations that have been made and future spectropolarimetric observations that will be made. These investigations have also clearly outlined the strengths and weaknesses of the photospheric model, hightlighting the role that global radiative transfer of realistically structured jets has in our understanding of GRBs. With this unprecedented knowledge, there is a clear path towards including different physics and further testing the radiation mechanisms that are believed to play a role in the GRB prompt emission. Furthermore, studies that produce testable and falsifiable spectropolarimetric predictions from other prompt emission models will help accelerate our understanding of this phase of GRBs, especially with rich datasets promised for the future.

\begin{acknowledgements} 
TP and DL acknowledge support by NASA grants 80NSSC18K1729 (Fermi) and NNX17AK42G (ATP), Chandra grant TM9-20002X, and NSF grant AST-1907955. TP acknowledges funding from the Future Investigators in NASA Earth and Space Science and Technology (FINESST) Fellowship, NASA grant 80NSSC19K1610. Resources supporting this work were provided by the NASA High-End Computing (HEC) Program through the NASA Advanced Supercomputing (NAS) Division at Ames Research Center. Additionally, this work used the CoSINe High Performance Computing cluster,  which is supported by the College of Science at Oregon State University.  
\end{acknowledgements}

\software{Astropy \citep{astropy:2013, astropy:2018}, Scipy \citep{Scipy}, Numpy \citep{Numpy}, MCRaT \citep{parsotan_mcrat_software_2021_4924630}, ProcessMCRaT \citep{parsotan_processmcrat_software_2021_4918108}}

\bibliography{references}
\listofchanges 

\end{document}